\documentclass[aip,jcp,amsmath,amssymb,reprint,floatfix]{revtex4-1}

\usepackage{graphicx}
\usepackage{dcolumn}
\usepackage{bm}
\relax
\newcommand{\Loss}{\mathcal{L}}
\newcommand{\fmode}{\mathcal{F}}
\newcommand{\Kcal}{\mathcal{K}}
\newcommand{\LIF}{^\text{LIF}}
\newcommand{\CRD}{^\text{CRD}}
\newcommand{\CELIF}{^\text{CELIF}}

\begin{document}
\title{Absolute absorption and fluorescence measurements over a dynamic range of 10$^\textbf{6}$ with cavity-enhanced laser-induced fluorescence}
 \author{Scott E. Sanders}
 \affiliation{Department of Chemistry, Durham University, South Road, Durham DH1 3LE, United Kingdom}
 \author{Oliver R. Willis}
 \affiliation{Department of Chemistry, Durham University, South Road, Durham DH1 3LE, United Kingdom}
 \author{N. Hendrik Nahler}
 \affiliation{School of Engineering and Physical Sciences, Heriot-Watt University, Edinburgh EH14 4AS, United Kingdom}
 \author{Eckart Wrede}
 \email{eckart.wrede@durham.ac.uk}
 \affiliation{Department of Chemistry, Durham University, South Road, Durham DH1 3LE, United Kingdom}%
 \date{\today}

\begin{abstract}
We describe a novel experimental setup that combines the advantages of both laser-induced fluorescence and cavity ring-down techniques. The simultaneous and correlated measurement of the ring-down and fluorescence signals yields absolute absorption coefficients for the fluorescence measurement. The combined measurement is conducted with the same sample in a single, pulsed laser beam. The fluorescence measurement extends the dynamic range of a stand-alone cavity ring-down setup from typically three to at least six orders of magnitude. The presence of the cavity improves the quality of the signal, in particular the signal-to-noise ratio. The methodology, dubbed cavity-enhanced laser-induced fluorescence (CELIF),  is developed and rigorously tested against the spectroscopy of 1,4-bis(phenylethynyl)benzene in a molecular beam and density measurements in a cell. We outline how the method can be utilised to determine absolute quantities: absorption cross sections, sample densities and fluorescence quantum yields.
 \end{abstract}

 \pacs{Valid PACS appear here}

 \maketitle

\section{Introduction}\label{sec:intro}

Laser-induced fluorescence (LIF) has become a well established spectroscopic technique since the advent of the laser.\cite{Donovan2007,Kinsey1977,Zare2012} It is an indirect absorption technique as the spontaneously emitted photons are recorded as signal after incident laser light has been absorbed by a species.  Most commonly, the wavelength integrated fluorescence is detected leading to a fluorescence excitation spectrum that is equivalent to an absorption spectrum.  Additional information can be obtained from the wavelength-dispersed fluorescence.
In gerneral, LIF possesses intrinsically low noise allowing the detection of very low concentrations in confined volumes. High spatial resolution is usually achieved by right-angle collection of the fluorescent light with a large-aperture lens. 
An absolute measurement of absorbance requires detailed knowledge of the fluorescence process and careful calibration of the detection system (fluorescence quantum yield, geometrical setup, spectral response of the detector, \emph{etc.}). 
Using amplified photodetectors, laser-induced fluorescence can be measured linearly over a large dynamic range from single-photon counting to saturation of the photodetector. The measurement is virtually background free if stray light from the incident laser is suppressed effectively, \emph{e.g.}\ with an optical filter. If the fluoresecence lifetime is long compared to the laser pulse length, stray light can also be discriminated against by time gating of the signal.
If the fluorescence lifetime exceeds the time between collisions, quenching of the fluorescence can occur, reducing the signal potentially to a level which makes a measurement impossible. In addition, quenching, predissociation and other non-radiative processes complicate a straightforward relationship between signal and concentration, due to possible dependences of these processes on the excited state. Nevertheless, LIF measurements have been carried out in different media from collision-free environments, \emph{e.g.}\ molecular beams, to liquids and solids.\cite{Brecher1976} 

Over the last two decades, cavity ring-down spectroscopy (CRDS) has become a well-established and widely applied spectroscopic technique.\cite{Berden2009,O'Keefe1988} CRDS is based on the Beer-Lambert law and performs a direct absorption measurement. Consequently, fluorescence of the sample is not a detection requirement. 
In a CRD experiment, light enters a cavity formed by two highly reflective mirrors (typically $R > 0.999$). During each pass a small fraction of light leaks out through the mirrors and is detected as an exponential decay behind the exit mirror. The inverse of the decay rate of the signal is referred to as ring-down time, which for an empty cavity is given by
\begin{equation}
  \tau_0 = \frac{d}{c (1 - R)},
\end{equation}
where $d$ is the distance between the mirrors of reflectivity $R$ and $c$ is the speed of light. Depending on the reflectivity of the mirrors, several thousand round trips can be achieved. With cavity lengths of the order of 1\,m the effective pathlengths are of the order of many kilometers. The absorption of light by a sample inside the cavity causes an additional loss and consequently a shorter ring-down time. The reduced ring-down time, $\tau$, is directly linked to the absorption coefficient
\begin{equation}\label{eq:alpha_CRD}
  \alpha = \sigma \rho = \frac{1}{c} \left(\frac{1}{\tau} - \frac{1}{\tau_0}\right),
\end{equation}
where $\sigma$ is the absorption cross section and $\rho$ is the number density of the sample, leading to a photon loss per pass of $\alpha d$. As the ring-down time is the measurable, the technique is immune to power fluctuations of the incident laser source leading to an absolute and self-calibrated measurement.

Eq.\ \ref{eq:alpha_CRD} assumes that the entire cavity is filled by the sample which is the case in a typical cell experiment. If the sample volume is localized, \emph{e.g.}\ inside a molecular beam, then the right-hand side of eq.\ \ref{eq:alpha_CRD} needs to be multiplied by the ratio of the cavity and sample lengths, $d/s$, to account for the higher density in the smaller sample volume.

In cavity ring-down spectroscopy, the absorbance is measured  over the pathlength through the sample (integrated column density) and, in contrast to LIF, is not spatially resolved. Similar to LIF, CRDS has been applied to gas-phase, liquid and solid samples. The direct absorption measurement removes the issue of fluorescence quenching but introduces the problem of large, unwanted losses in liquid and solid samples (scattering and reflection). CRDS measurements are not background free as they are based on the detection of a change in signal. The dynamic range is defined by the minimal and maximal detectable change in ring-down time. In a typical CRDS setup, the dynamic range spans three orders of magnitude.\cite{Lehmann2008} The sensitivity of LIF is generally superior for spatially confined samples whereas CRDS can match or exceed this sensitivity for larger samples that fill the entire length of the cavity.

In this study, we present a novel, direct combination of the CRD and LIF techniques using a single, pulsed laser beam that we name cavity-enhanced laser-induced fluorescence or CELIF. The CRD and LIF signals are recorded simultaneously on a shot-to-shot basis. For the first time the fluorescence signal, cavity ring-down intensity and cavity ring-down time are cross-correlated in a way that significantly enhances both techniques. The CRD measurement provides the absolute calibration of the LIF signal. The combined techniques lower the detection limit of the CRD measurement by several orders of magnitude.
 
Previously, several groups have used different combinations of CRD and LIF, not necessarily using a single laser beam, to measure, \emph{e.g.}, fluorescence quantum yields and quenching rates.\cite{Spaanjaars1997,Hagemeister1999,Bahrini2006,Tokaryk2007} Richman \emph{et al.}\ detected fluorophor-doped aerosols within a cavity by their fluorescence signal.\cite{Richman2005} Furthermore, CRD spectroscopy, instead of Rayleigh scattering, has recently been used to calibrate density measurements in flames via LIF.\cite{Dreyer2001,Luque2004,Lamoureux2010} None of these previous studies used the cross-correlation of the CRD and LIF signals as presented here.

In this paper, we describe and fully characterize the novel CELIF technique in detail. We outline how absolute quantities, such as fluorescence quantum yields, can be directly extracted from the single beam CELIF measurement by cross-normalization of the LIF and CRD signals. Absorption coefficients are accessible with a single method and measurements can be carried out over a dynamic range spanning more than six orders of magnitude. The ring-down cavity rejects the majority of the laser light and stretches the laser pulse in time. The amount of light inside the cavity, generating the LIF signal, is measured by the integrated cavity ring-down signal. This measurement is used for a very robust shot-to-shot normalization of the LIF signal against the light intensity, leading to a much enhanced signal-to-noise ratio. In comparison to a single-pass LIF measurement, CELIF greatly reduces saturation and power broadening due to the much lower photon densities and effectively eliminates stray light due to the transversal mode structure of the cavity. 

Section \ref{sec:experiment} describes the methodology and how the integrated CRD signal is used for the shot-to-shot normalization of the LIF signal with respect to the fluctuating intensity of a pulsed laser. Absolute absorption coefficients determined from the CRD measurement can then be used to cross calibrate the LIF signal. In this work, CELIF was implemented using a pulsed molecular beam setup and a cell experiment. Section \ref{sec:results} presents measurements that scrutinize the methodology, particularly the increased dynamic range. In section \ref{sec:discussion}, we discuss the characteristics of the method including its enhanced limit of detection and how absolute fluorescence quantum yields can be determined. In the conclusions, we outline the power of the CELIF technique and its most suitable applications.

\section{Experiment}\label{sec:experiment}

The fundamental idea of this setup is the combination of CRDS and LIF in a single, pulsed laser beam. A schematic layout of the setup is shown in fig.\ \ref{fig:exp}. It is a straightforward combination of a classical LIF and pulsed CRDS setup: the sample is intersected by the laser beam that is confined in the cavity and the laser-induced fluorescence is collected at right angles. In our setup the wavelength-integrated fluorescence is recorded. Without loss of generality, the technique can be used in a setup where the dispersed fluorescence spectrum is measured. In principle, the technique should be widely applicable to any gas-phase, liquid and even solid samples which have been used in previous CRDS studies as long as the fluorescence light can be extracted from the sample volume.

In the following subsection we derive how in our combined CRD/LIF setup the cavity ring-down signal is used to normalize and absolutely calibrate the LIF signal.

\subsection{CELIF methodology}
\label{sec:method}

In a general LIF measurement, the time-integrated LIF signal, $S\LIF$, is proportional to the light intensity, $I\LIF$, that has interacted with the sample within the LIF probe volume:
\begin{equation}\label{eq:SLIF}
	S\LIF(\lambda) = \alpha(\lambda) \cdot \Gamma(\lambda) \cdot g \cdot I\LIF,
\end{equation}
where $\lambda$ is the excitation wavelength, $\Gamma$ is the fluorescence quantum yield and $g$ is a geometry dependent factor of the detection system. In principle, $g$ is also a function of $\lambda$ as the fluorescence spectrum may depend on the excited state. However, this dependence can only be considered if the dispersed fluorescence spectrum is recorded as a function of excitation wavelength. In order to obtain the absorption coefficient, $\alpha$, from a fluorescence excitation measurement, $S\LIF$ and $I\LIF$ need to be measured. The factor $g$ is an instrument function that is not readily available but can be determined via a meticulous external calibration.  The fluorescence quantum yield, $\Gamma$, is generally unknown and needs to be measured or predicted from theory in a separate study.

Fundamentally, CELIF is a LIF measurement where the simultaneous CRD measurement is used for the normalization and the calibration of the LIF signal.  In the following, we derive how the time-integrated CRD signal is correlated to $I\LIF$ and how it is subsequently used to provide the normalization of $S\LIF$ to eliminate shot-to-shot fluctuations in the laser intensity from eq.\ \ref{eq:SLIF}.  We then describe how the absolute absorption coefficient determined via the ring-down time measurement (eq.\ \ref{eq:alpha_CRD}) is used for the absolute calibration of the normalized LIF signal such that in a CELIF measurment prior knowledge of $g$ and $\Gamma$ is not required.
 
However, there are differences in the measurement of the fluorescence signal between a CELIF and a typical LIF experiment. Only a fraction of the laser light is transmitted through the cavity entrance mirror and is subsequently interacting with the sample. As a consequence, the initial fluorescence is very small in comparison to single-pass LIF. The light confined in the cavity undergoes up to several thousand round trips, interacting with the sample twice on each of them. Each of these interactions induces further fluorescence so that the integrated fluorescence leads to an appreciable LIF signal. Effectively, the cavity stretches the short laser pulse into the exponential decay characteristic for the cavity.

Without loss of generality, we base the following derivation on a fixed excitation wavelength and on a sample distribution that is symmetric with respect to a LIF probe volume that is placed at the center of the cavity. The photon loss per single pass is $\Loss = \sigma \rho s $, with the sample length, $s$, given that $\Loss \ll 1$. The light intensity at the center after entering the cavity is
\begin{equation}\label{eq:ILIF0}
I_0\LIF = I_\text{L} T \fmode \left( 1 - \frac{\Loss}{2}\right),
\end{equation}   
where $I_\text{L}$ is the laser intensity incident on the cavity entrance mirror and $T\fmode$ is the fraction of light that can be resonantly coupled into the cavity.  $T$ is the transmission of the mirror and $\fmode$ is the fraction of the laser bandwidth that is resonant with the mode structure of the cavity.

In addition to the sample loss, a fraction of light leaks out of the cavity upon reflection at the mirror. The light intensity at the center of the cavity after $i$ single passes is
\begin{equation}
I_i\LIF = I_0\LIF \left[ \left(1-\Loss\right) R\right]^i,
\end{equation}
where $R$ is the reflectivity of the cavity mirrors. Using the summation rule for geometric series, the summed light intensity, $I_n\LIF$, that has crossed the LIF probe volume after $n$ single passes and the integrated intensity in the limit $n\to\infty$, $I\LIF$,  are:
\begin{align}
I_n\LIF &= I_0\LIF \sum_{i=0}^n \left[ \left(1-\Loss\right) R\right]^i \nonumber\\
	&= I_0\LIF \frac{{1 - \left[(1 -\Loss) R\right]^{n+1}}}{1 - (1 -\Loss) R}\\
\label{eq:ILIFint}
I\LIF &= \lim_{n\to\infty} I_n\LIF = \frac{I_0\LIF}{1 - (1 -\Loss) R}
\end{align}
Using eq.\ \ref{eq:ILIF0} and the approximations $\mathcal{L}\ll 1$ and $T\approx 1 - R$ shows that $I\LIF\approx I_\text{L}\mathcal{F}$. This means that, through the repeated use of the decaying light pulse inside the cavity, the amount of light that interacts with the sample is the amount incident on the entrance mirror that is resonant with the mode structure of the cavity.  We will discuss the value of $\fmode$ for our setup in section~\ref{sec:CELIF_characteristics}.

Similarly, we can derive the integrated CRD intensity, $I\CRD$, at the exit mirror taking into account that, following the initial single pass, light is only collected after every round trip:
\begin{align}
I_0\CRD &= I_\text{L} T^2 \fmode \left( 1 - \Loss\right)\label{eq:ICRD0}\\
I_n\CRD &= I_0\CRD \sum_{j=0}^{n/2} \left[ \left(1-\Loss\right) R\right]^{2j} \nonumber \\
	&= I_0\CRD \frac{{1 - \left[(1 -\Loss)^2 R^2\right]^{n+1}}}{1 - (1 -\Loss)^2 R^2}\\
I\CRD &= \lim_{n\to\infty} I_n\CRD = \frac{I_0\CRD}{1 - (1 -\Loss)^2 R^2}\label{eq:ICRD}
\end{align}

In the limit of an empty cavity ($\Loss = 0$) and $T=1-R$, the measured, time-integrated light intensity imparting on the detector is
\begin{equation}\label{eq:transmission}
I\CRD = I_\text{L} T\fmode/2
\end{equation}
where $T\fmode/2$ is the cavity transmission function. The comparison of the integrated LIF and CRD intensities yields:
\begin{equation}\label{eq:ILIF_accurate}
I\LIF = I\CRD\frac{[1+(1-\Loss)R](1-\Loss/2)}{T(1-\Loss)}
\end{equation}
In the limit of small loss $\Loss \ll 1$ and additionally $R \approx 1$, $I\LIF$ can be approximated as
\begin{equation}\label{eq:ILIF_approx}
I\LIF \approx I\CRD\frac{1+R}{T} \approx I\CRD\frac{2}{T}
\end{equation}
As an example, for a relatively poor mirror reflectivity of $R = 0.998$ and a large loss of $\Loss = 0.001$, which would be amenable to a single-pass absorption measurement, the relative error of this approximation is smaller than $10^{-3}$. This error is dominated by the mirror reflectivity and reduces to $5\cdot 10^{-5}$ for $R = 0.9999$.

Therefore, we define the normalized CELIF signal as
\begin{equation}\label{eq:CELIF_norm}
S\CELIF = \frac{S\LIF}{I\CRD}.
\end{equation}
Note that $S\CELIF$ is a relative quantity (unitless in our case) that can be calibrated to equal the absolute absorption coefficient, $\alpha$, derived from the measurement of ring-down times.  The calibrated CELIF absorption coefficient is
\begin{equation}\label{eq:alpha_LIF}
\alpha = \sigma\rho = \Kcal\cdot S\CELIF,
\end{equation}
where $\Kcal$ is the proportionality factor that can be determined from a simultaneous LIF and CRD measurement, provided the absorption leads to a sufficient reduction in ring-down time (equations \ref{eq:alpha_CRD} and \ref{eq:alpha_LIF}):
\begin{equation}\label{eq:cal_LIF}
\Kcal = \frac{I\CRD}{S\LIF} \frac{1}{c} \left(\frac{1}{\tau} - \frac{1}{\tau_0}\right).
\end{equation}

This method provides an absolute calibration of the LIF signal. Therefore the limited dynamic range of the CRD method can  be extended towards the generally lower detection limit of the LIF method. The combination of LIF and CRDS into CELIF enables the measurement of an absolute absorption coefficient over the combined dynamic range.

Equations \ref{eq:alpha_LIF} and \ref{eq:cal_LIF} hold true for an absorption measurement at a fixed wavelength. Generally, $\Kcal$ is a function of the excitation wavelength (fluorescence quantum yield, mirror transmission) and the fluorescence spectrum (spectral response of the detection system). A single-pass LIF measurement needs to account for the wavelength dependencies of the quantum yield and the detection system. In addition, a calibrated CELIF spectrum requires knowledge of the wavelength-dependent exit mirror transmission. In section \ref{sec:yield} we describe a procedure that uses the simultaneously measured ring-down times and Rayleigh scattering collected with the LIF detection setup to determine the wavelength dependence of $T$ based on our unchanged setup. With two sets of UV and visible mirrors we found the transmission to vary by less than 5\% over a wavelength range of 1~nm which will allow to approximate the transmission as constant in many applications. We note, however, that the common approximation $T = 1 - R$, with $R(\lambda)$ derived from an empty-cavity ring-down scan, could not be generally applied over the usable wavelength range of each set of mirrors.

\subsection{Apparatus}

\begin{figure}
  \includegraphics[width=8.6cm]{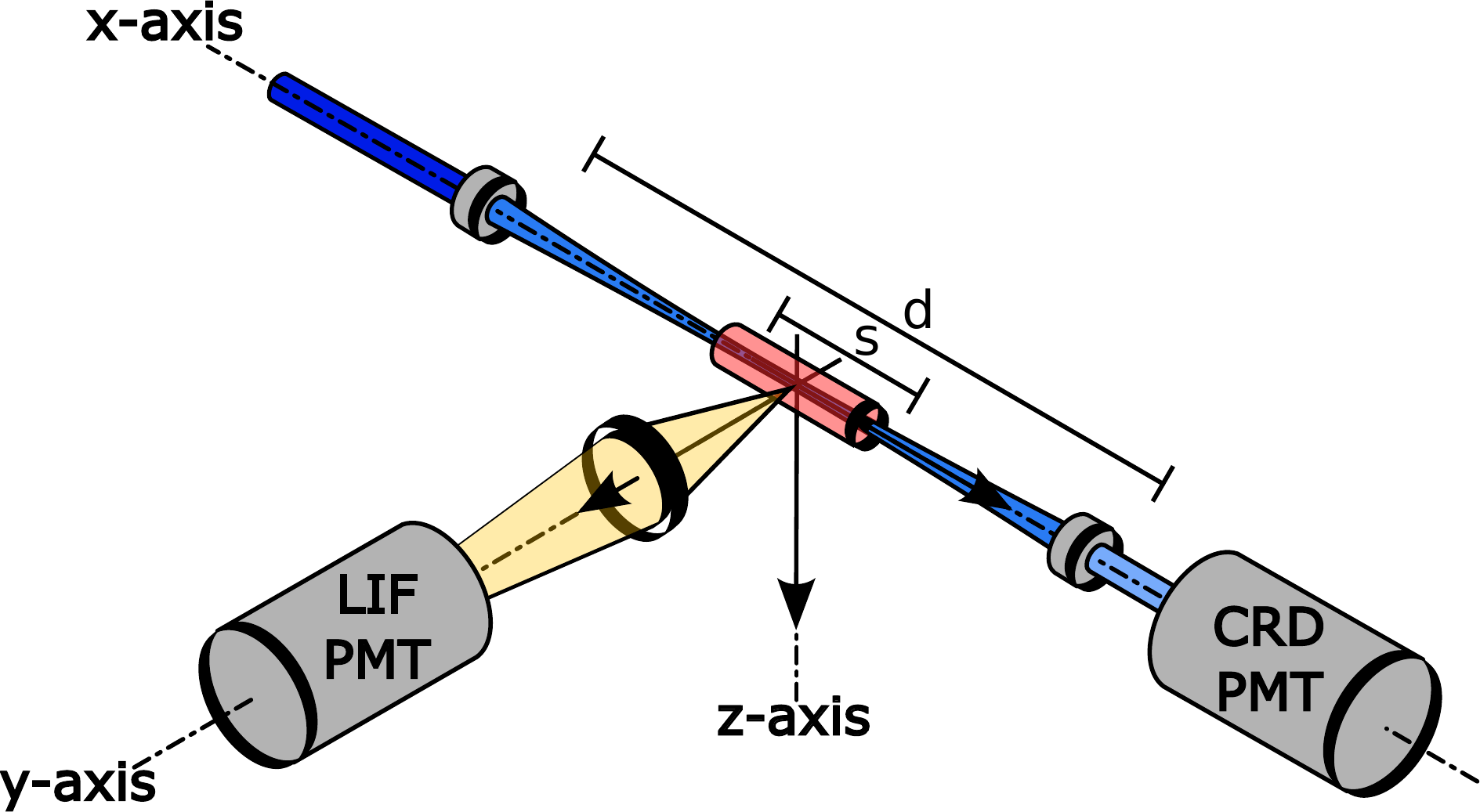}
\caption{Experimental setup. The sample volume is situated at the center of the ring-down cavity and the fluorescence is collected at right angles.}
\label{fig:exp}
\end{figure}

A general CELIF apparatus is shown in fig.\ \ref{fig:exp}. It consists of a standard ring-down cavity including beam-shaping, mode-matching optics and a suitably fast photodetector (x-axis in figure). A typical LIF detection system, including a collimation lens and photodetector, is added at right angles to the cavity axis, preferably at the center of the sample (y-axis in figure).  Two experimental setups were used in this study. Setup~1 introduced the sample via an unskimmed molecular beam (z-axis in figure) whereas setup~2 was used for cell measurements where the entire cavity was filled with the sample gas. 

Setup~1 was based on our CRD spectrometer that was used to study the torsional motions of jet-cooled 1,4-bis(phenylethynyl)benzene (BPEB), the details of which are described in ref.\ \onlinecite{Greaves2006}. Briefly, solid BPEB was sublimated in a heated oven attached to the front of a pulsed solenoid valve (Parker, General Valve Series 9). The gaseous BPEB was picked-up by the argon carrier gas in the channel of the oven and cooled in the subsequent supersonic expansion. The BPEB sample density could be controlled over more than three orders of magnitude by varying the oven temperature. The cavity axis crossed the molecular beam approximately 5~mm downstream of the oven orifice. The $\text{S}_1 \leftarrow\text{S}_0$ transition of BPEB is very strong\cite{Beeby2002,CrossSectionNote} and was excited over a wavelength range of 317--321.5~nm. Over this range, the reflectivity of the cavity mirrors (Layertec, center wavelength 330~nm) varied from 99.8 to 99.9\% leading to empty-cavity ring-down times of $1.2-2.5~\mu$s (cavity length $\approx 84$~cm). The doubled output of the dye laser (Sirah Cobra-Stretch, pumped by Continuum Surelite I-10, 10~Hz repetition rate) ranged from 30 to $100~\mu$J with a bandwidth of 0.045~cm$^{-1}$ and a pulse length of $\approx 5$~ns. The LIF collection optics imaged the full probe volume (overlap of molecular and laser beams) onto the photodetector.

With setup~2 we carried out N$_2$ Rayleigh scattering (at 583.5~nm) and acetone fluorescence (at 313~nm) measurements to verify the methodology as described in section\ \ref{sec:method}. We note that the same detector setup was used for the Rayleigh scattering and fluorescence measurements. In general, CELIF signal is referred to signal collected by the LIF detection setup normalized by the CRD intensity. Therefore CELIF signal in the course of this paper refers to Rayleigh scattering or fluorescence signals. The gas pressure of the filled cavity (length $\approx 81$~cm) was monitored over a range of $0.01-1000$~mbar with capacitance manometers (Leybold CR090/CTR100 and MKS 626A) to cross reference the spectroscopically determined sample densities. For the Rayleigh scattering measurements, empty-cavity ring-down times in excess of 40~$\mu$s were achieved with the corresponding set of mirrors ($R>99.99\%$). For the fluorescence measurements, the same set of mirrors was used as in setup~1 and the drop-off in their reflectivity to about 99.7\% shortened the ring-down time to $\approx 800$~ns. The output of the (doubled) dye laser (Quanta-Ray PDL-2, pumped by Continuum Minilite II, 10~Hz repetition rate) was $\approx 300~\mu$J at 583.5~nm and $\approx 100~\mu$J at 313~nm, both with a bandwidth of 0.3~cm$^{-1}$. The LIF optics imaged a 3~mm long probe volume onto the photodetector.

In both setups we use two identical photomultipliers (Hamamatsu, H7732-10 module with R928 tube), the signals of which were simultaneously recorded using a two-channel digitizer (National Instruments NI PCI-5124, 12-bit for setup 1 and AlazarTech ATS460, 14-bit for setup 2). 

\subsection{Data acquisition and analysis}\label{sec:daq}

For each laser shot, the LIF transient and the CRD transient are measured simultaneously as shown in fig.\ \ref{fig:transients} (a) and (b), respectively. The CRD transient follows the typical exponential decay (fig.\ \ref{fig:transients} (b)) from which the ring-down time, $\tau$, is extracted by a non-linear least squares fit. In combination with the empty-cavity ring-down time, $\tau_0$, the absorption coefficient, $\alpha$, is determined as in a typical CRD experiment. In the following text, a CRD measurement or spectrum is based on the ring-down times according to eq.\ \ref{eq:alpha_CRD}.

We also integrate the CRD transient to extract $I\CRD$ (\emph{cf.} eq.\ \ref{eq:ICRD}) which is proportional to the light intensity inside the ring-down cavity (eq.\ \ref{eq:ILIF_approx}). The LIF transient which follows the ring-down decay is integrated to yield the LIF signal, $S\LIF$. According to eq.\ \ref{eq:CELIF_norm} the CELIF signal, $S\CELIF$, is obtained through a shot-to-shot normalization of $S\LIF$ with respect to $I\CRD$.

As alluded to earlier, CELIF is a LIF measurement where, without further calibration, only relative quantities are obtained. In CELIF this absolute calibration is provided by the absorption coefficient, $\alpha$, from the simultaneous CRD measurement, see eq.\ \ref{eq:alpha_LIF}. This calibration is particularly robust as both the LIF and CRD measurements use the same laser photons and sample molecules.

\section{Results}\label{sec:results}

\begin{figure}
  \includegraphics[width=8.6cm]{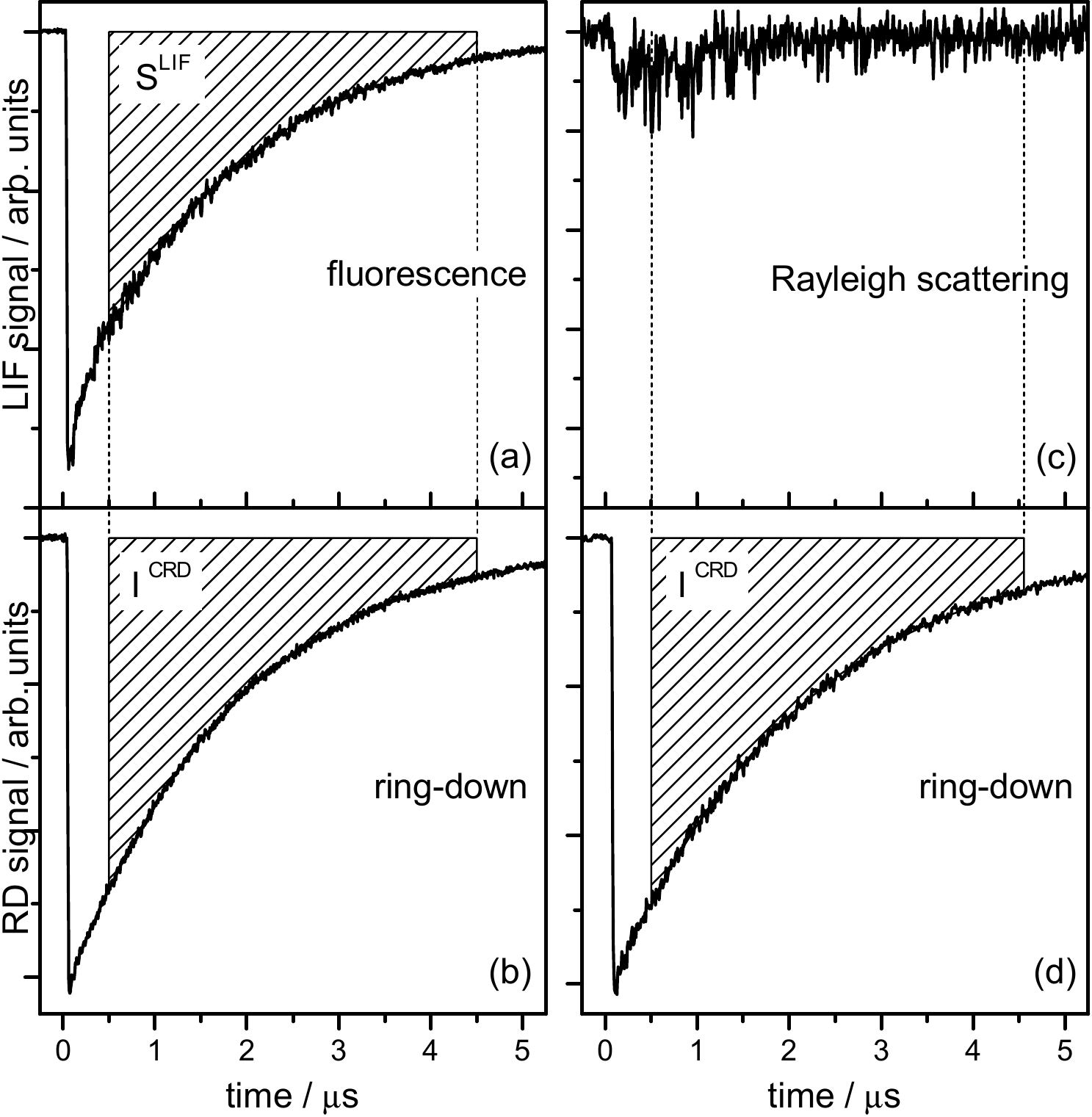}\\
\caption{Simultaneously recorded (a) LIF and (b) CRD transients of the transition at 319.44~nm (\emph{cf.}\ fig.\ \ref{fig:BPEBspectra}) of BPEB. The LIF transient follows the same exponential decay as the CRD transient due to the negligible sub-ns fluorescence lifetime of the excited state\cite{Fujiwara2008} compared to the 1.80~$\mu$s ring-down time. The same integration limits were used to derive $S\CELIF$, see eq.\ \ref{eq:CELIF_norm}. Simultaneously recorded  (c) Rayleigh scattering (measured with the LIF detector) and (d) CRD transients of BPEB at 321.0~nm.
The vertical scale in panel (c) is magnified to show the presence of Rayleigh scattering. The absolute noise is equivalent to the baseline noise in panel (a).}
\label{fig:transients}
\end{figure}

Based on our previous CRD work,\cite{Greaves2006} we chose BPEB as our model system.  One of its characteristic properties is a fluorescence lifetime of $<1$~ns following the $\text{S}_1 \leftarrow \text{S}_0$ electronic excitation.\cite{Fujiwara2008} The left column of fig.\ \ref{fig:transients} shows samples of simultaneously recorded LIF and CRD transients on resonance.  As the fluorescence lifetime is negligible compared to the 1\,--\,2~$\mu$s ring-down times, the LIF transient follows the exponential decay of the ring-down signal. Based on this fact, we chose the same integration boundaries for the LIF signal ($S\LIF$) and ring-down intensity ($I\CRD$), indicated by the hashed areas in the figure, to derive the normalized CELIF signal ($S\CELIF$). In the more general case, where the fluorescence lifetime is no longer small in comparison to the ring-down time, both the entire LIF and CRD transients need to be integrated to ensure correct normalization.

\begin{figure}
  \includegraphics[width=8.6cm]{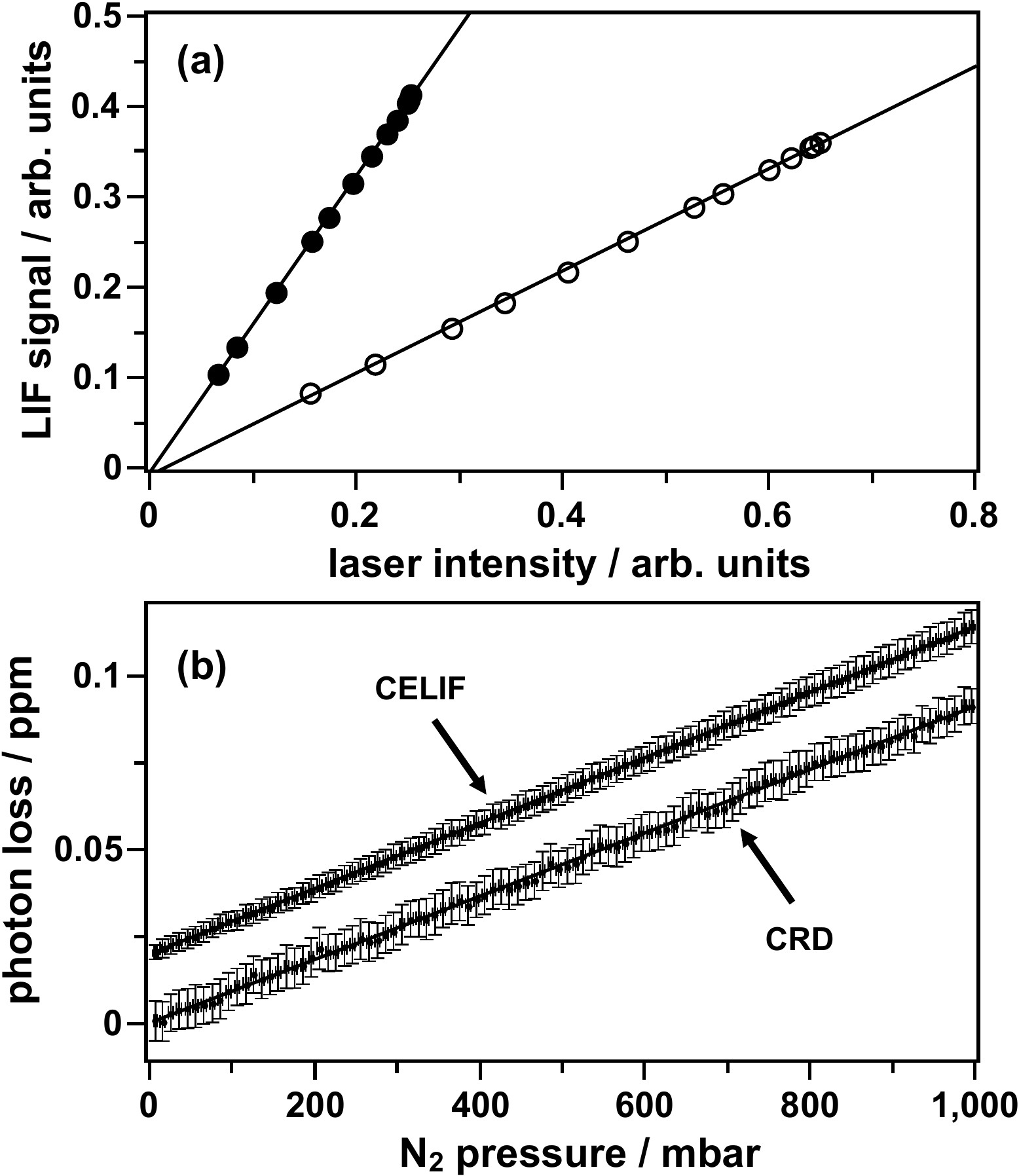}\\
\caption{(a) Dependence of the LIF signal ($S\LIF$) on the incident laser intensity as measured by the ring-down detector ($I\CRD$). Data has been recorded using the fluorescence of acetone at the excitation wavelength of 313~nm at two pressures of 0.1\ mbar ($\circ$) and 0.3\ mbar ($\bullet$) leading to losses per pass of $(1.42\pm 0.02)\cdot 10^{-3}$ and $(4.31\pm 0.04)\cdot 10^{-3}$ respectively (linear least-squared fit). Error bars are significantly smaller than the symbol size. (b) Pressure-dependent N$_2$ Rayleigh scattering at 583.5~nm. The photon loss is derived from the ring-down times of CRD measurement (eq.\ \ref{eq:alpha_CRD}) and the CELIF data (shifted for clarity) is calibrated according to eq.\ \ref{eq:alpha_LIF} at 1~bar. Each data point is based on an average of 311 laser shots in a 10~mbar pressure range. The error bars represent the spread of the data.
The solid lines are the linear least-squared fits to the data.}
\label{fig:powerdep}
\end{figure}

In order to assess detection limits and signal-to-noise ratios, we also measured off-resonance transients of BPEB at 321~nm shown in the right column of fig.\ \ref{fig:transients}. As BPEB has an extended $\pi$-system along its major axis leading to a large polarizability, we can detect its Rayleigh scattering in a seeded molecular beam (fig.\ \ref{fig:transients}c) with our CELIF setup. The corresponding CRD transient (fig.\ \ref{fig:transients}d) did not cause a reduction in ring-down time with respect to the empty-cavity demonstrating the higher sensitivity of the CELIF setup.

To confirm the validity of eqs \ref{eq:ILIF_accurate} to \ref{eq:alpha_LIF}, we measured the linearity of the LIF signal with respect to the laser intensity, as measured via the integrated CRD intensity, $I\CRD$, and the dependence of the CELIF signal, $S\CELIF$, on the sample density, $\rho$. 
Fig.\ \ref{fig:powerdep}(a) shows the fluorescence signal of acetone following laser excitation at 313~nm as a function of laser intensity. The cavity was filled with 0.1 and 0.3~mbar of acetone respectively. Each data point represents the average over 5000 laser shots. The standard errors in laser intensity and LIF signal are smaller than the size of the symbols. The linear least-squares fits demonstrate the validity of the shot-to-shot normalization using eq.\ \ref{eq:CELIF_norm} and show the absence of any saturation effects in the measurement range.

The linearity of the CELIF signal with respect to the sample density of a filled cavity was confirmed with N$_2$ Rayleigh scattering at 583.5~nm (fig.\ \ref{fig:powerdep}b) in the pressure range 0.1\,--\,1000~mbar. The simultaneously measured ring-down times  provide an absolute measurement for the absorption and therefore the photon loss per pass using eq.\ \ref{eq:alpha_CRD}. The Rayleigh scattering signal obtained by the CELIF setup was calibrated to match the CRD photon loss at a N$_2$ pressure of 1~bar according to eq.\ \ref{eq:alpha_LIF}. The slopes of the independent linear least squared fits to the data in fig.\ \ref{fig:powerdep}b are indistinguishable proving the validity of eqs\ \ref{eq:CELIF_norm}--\ref{eq:cal_LIF}. 

Moreover, we found that during these experiments with variable cavity pressures, movement of the mirrors could be observed by the ring-down measurement. Only after we improved the mechanical stability of our cavity and mirror mounts, we were able to set up a cavity unaffected by the change from vacuum to atmospheric pressure. This was verified by comparison of the CRD density measurement to the reading of the capacitance manometer. However, the simultaneous CELIF measurement is immune to these cavity misalignments as these are fully compensated for by the shot-to-shot normalization procedure leading to higher quality data.

\begin{figure}
  \includegraphics[width=8.6cm]{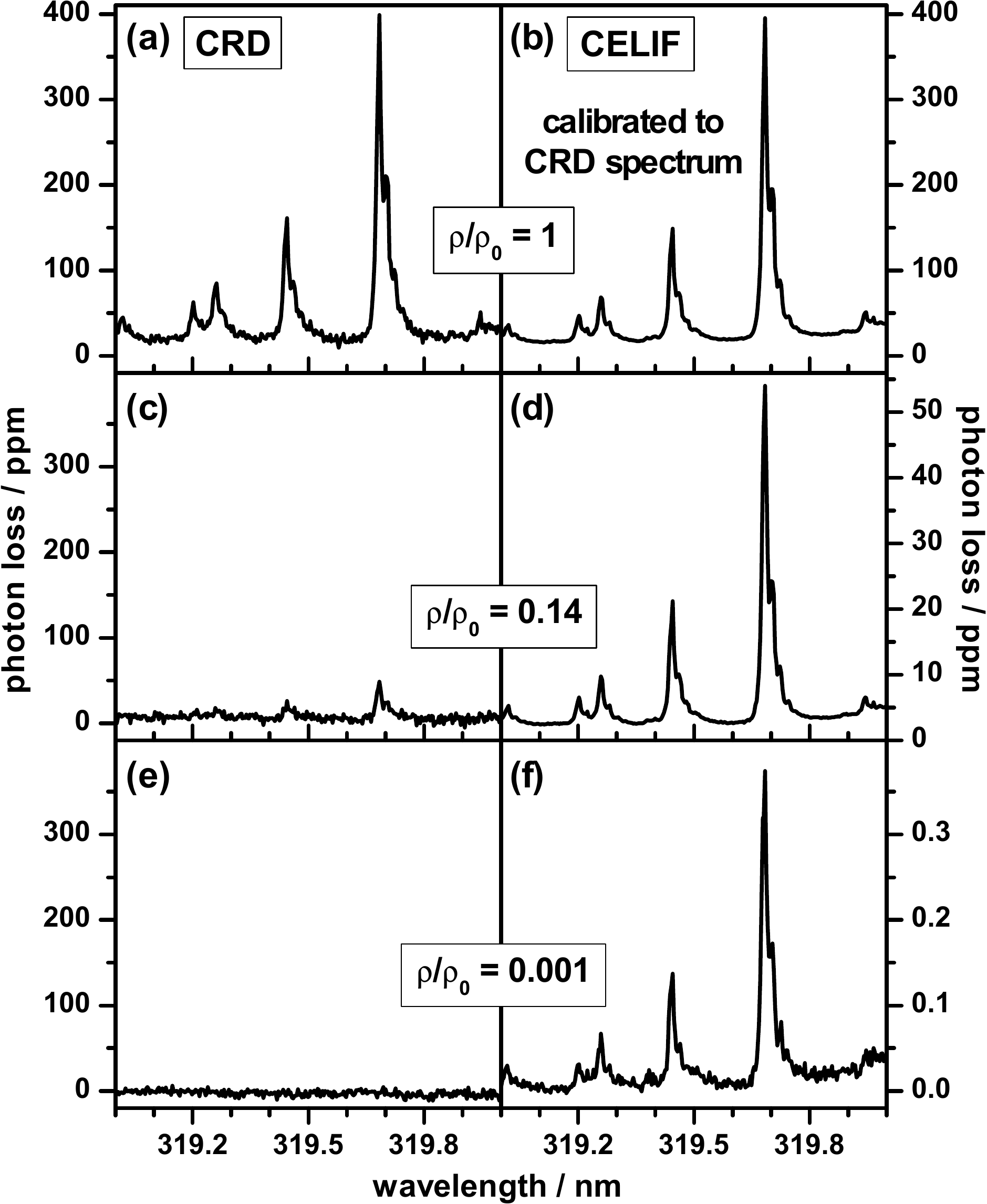}
 \caption{Comparison of simultaneously recorded CRD (left column) and CELIF (right column) spectra of jet-cooled BPEB at different relative concentrations, $\rho/\rho_0$. The CRD spectra are shown to the same scale as the absolute noise does not change with concentration. The absolute photon loss per pass of the CELIF spectrum (b) was calibrated to the simultaneously recorded CRD spectrum (a). The CELIF spectra (d) and (f) were calibrated with respect to spectrum (b). For details of the BPEB spectroscopy see ref.~\onlinecite{Greaves2006}.}
\label{fig:BPEBspectra}
\end{figure}

Figure \ref{fig:BPEBspectra} shows a series of simultaneously recorded CRD and CELIF spectra of jet-cooled BPEB as a function of relative sample density $\rho / \rho_0$. The dynamic range of this particular CRD experiment is rather poor as shown in the left column. The main limitations are set by the small sample volume (diameter of the molecular beam), the shot-to-shot variation of sample concentration and the comparatively low reflectivity of the ring-down mirrors. In contrast, the CELIF measurement covers a much larger dynamic range as shown in the right column. With a change in sample density, the amplification of the photodetector was adapted to avoid saturation of the detection system. However, the series of CELIF measurements was internally calibrated each time the amplification of the photodetector was changed and we were able to follow the relative sample density, $\rho / \rho_0$, over a range of more than three orders of magnitude. The very low baseline noise in the CELIF measurements is evident in panels (b) and (d). Even in panel (f) where the sample density is reduced by three orders of magnitude the CELIF baseline noise is comparable to the CRD baseline noise at the highest sample concentration, panel (a).

\section{Discussion}\label{sec:discussion}

Our work has shown that recorded CELIF signals are linearly related to absorbances derived from CRD measurements. In this single laser beam setup, the cavity ring-down part provides the absolute scale for the LIF measurement without the requirement of any external calibration. With this combination of the two techniques, the high sensitivity of LIF greatly extends the accessible absorbance measurement range of CRDS on an absolute scale due to a greatly improved signal-to-noise ratio. The presence of the cavity introduces changes to the LIF technique that we will discuss in the following sections.

\subsection{Sources of noise and error}

Fig.\ \ref{fig:BPEBspectra} clearly shows the improvement in quality of the CELIF spectra compared to the CRD spectra. We will first discuss the sources of noise (or error) in both techniques. Shot-to-shot fluctuations in the sample density affect both techniques similarly as the same molecules are probed. 

For the CRD technique, mode fluctuations inside the cavity lead to increased noise on the ring-down transient, which, together with the electronic noise of the detector and the digitization error, increases the error in the exponential fit. All these sources of noise/error contribute to the shot-to-shot fluctuations in the measured ring-down time and hence the absorption coefficient, $\alpha$. A changing mirror alignment, \emph{e.g.}\ due to pressure changes inside the cell, will lead to a systematic error in $\alpha$, as we observed in our early measurements of the pressure dependence of the N$_2$ Rayleigh scattering (\emph{cf.}\ fig.\ \ref{fig:powerdep} (b)). Saturation of the photodetector (non-linear amplification) may also introduce a systematic error as the ring-down transient may be distorted. Although, this can easily be avoided by adjusting the CRD signal level via the laser pulse energy and the PMT voltage. 

In contrast, mode fluctuations or poor mirror alignment do not affect the CELIF measurement as only the integrated LIF and CRD signals, $S\LIF$ and $I\CRD$ respectively, are analyzed and not the shape of their transients. Thus, the noise/error in the integrals is only due to the noise of the detectors and the digitization errors. Saturation of the photodetectors needs to be avoided as described above. The strength of the CELIF technique is the precise determination of the laser intensity in the cavity, $I\LIF$, via the integrated ring-down intensity, $I\CRD$. The subsequent normalization, $S\LIF/I\CRD$, minimizes the effect of laser shot-to-shot fluctuations on $S\CELIF$. Overall, the noise in $S\LIF$ and $I\CRD$ is much lower than the noise/error in the determination of the ring-down time leading to the observed, low baseline noise in the CELIF spectra.

\subsection{General characteristics of CELIF measurements}\label{sec:CELIF_characteristics}

The cavity invokes stringent conditions on the shape and spectral composition of the beam interacting with the sample. Well designed and aligned cavity setups can be excited in a single transversal mode, TEM$_{00}$, leading to a well defined Gaussian beam waist at the center. This confinement effectively eliminates any stray light from an empty cavity, greatly reducing the background signal of the CELIF measurement. In contrast to a single-pass LIF setup, baffles and optical filters to suppress stray light are not necessary. This allows the detection of the whole fluorescence spectrum, including fluorescence on the excitation wavelength. As we have demonstrated in fig.\ \ref{fig:powerdep}, very clean measurements of Rayleigh scattering are possible. 

The mirror reflectivity and cavity length define the longitudinal mode structure supported by the cavity. The effect of the mode structure on a CRD measurement has been discussed in detail by Zalicki and Zare\cite{Zalicki1995} and needs to be taken into account similarly for CELIF. If we briefly consider a mode-matched cw cavity setup, the laser intensity inside the cavity will build up until the equilibrium is reached. This intensity (resonant with a cavity mode) is by a factor of the finesse higher than the incident laser intensity.  However, in our setups we use non-fourier-limited laser pulses, the spectral bandwidths of which span in the order of ten cavity modes. As the laser pulse length is comparable to the round-trip time, the cavity mode structure is not fully formed. The cavity transmission function, $T\fmode/2$ (\emph{cf.}\ eq.\ \ref{eq:transmission}) of a comparable case using a square temporal laser pulse is discussed in Ref.\ \onlinecite{Zalicki1995}, from which we conclude that $0.5 < \fmode < 1$ for our cavities. 

As outlined in section \ref{sec:method}, through the repeated use of the light pulse inside the cavity, the integrated light intensity creating the LIF signal in a CELIF experiment is $I_\text{L}\fmode$. In a single-pass, pulsed LIF experiment the fluorescence signal is created by the full intensity of the probe laser, $I_\text{L}$. Therefore, the total light intensity in both techniques differs only by the factor $T\fmode/(1-R)$, \emph{cf.}\ eqs \ref{eq:ILIF0} and \ref{eq:ILIFint}. Note that $\fmode$ cancels in the normalization of the LIF signal by the integrated CRD intensity in eq.\ \ref{eq:ILIF_accurate}. In terms of number of photons interacting with the sample, CELIF and single-pass, pulsed LIF are comparable. However, the photon flux per sample pass is several orders of magnitude lower for CELIF resulting in much reduced power broadening of spectral lines. 

The amount of molecules that is excited in any given time interval is proportional to the number of laser photons present in the cavity. Consequently, the time evolution of the LIF signal is the temporal convolution of the ring-down and fluorescence decays. The presence of the cavity effectively stretches the initial laser pulse in time, as characterized by the ring-down decay. Thus, like in a CRDS experiment, the ring-down time defines the temporal resolution of the CELIF setup with which the sample evolution can be monitored.  Brown \emph{et al.}\ have shown how, in a CRD experiment, the time evolution of the sample density can be extracted from a (non-exponential) ring-down decay using a forward convolution.\cite{Brown2000a} If sample densities change on the timescale of the ring-down time, similar techniques can be applied to map the temporal evolution of the sample density using CELIF.

\subsection{Limit of detection and extension of dynamic range}

The dynamic range of absorbance measurements is at one end determined by the limit of detection (LOD) defined by the noise level and at the other end by saturation. It is important to note that in CRD measurements the noise level stays almost constant across the entire dynamic range, see fig.\ \ref{fig:powerdep}b. The lowest absorbance that can be measured needs to cause a signal change on the ring-down trace greater than the overall noise on the trace. At the other end, CRD measurements are not valid any more when large absorbances lead to very short ring-down decay times. In a very carefully set up pulsed CRD measurement, the dynamic range in absorbance may cover three orders of magnitude. A typical pulsed CRD measurement in the UV spectral range spans two orders of magnitude in dynamic range.\cite{Wang2000}

\begin{figure}
  \includegraphics[width=8.6cm]{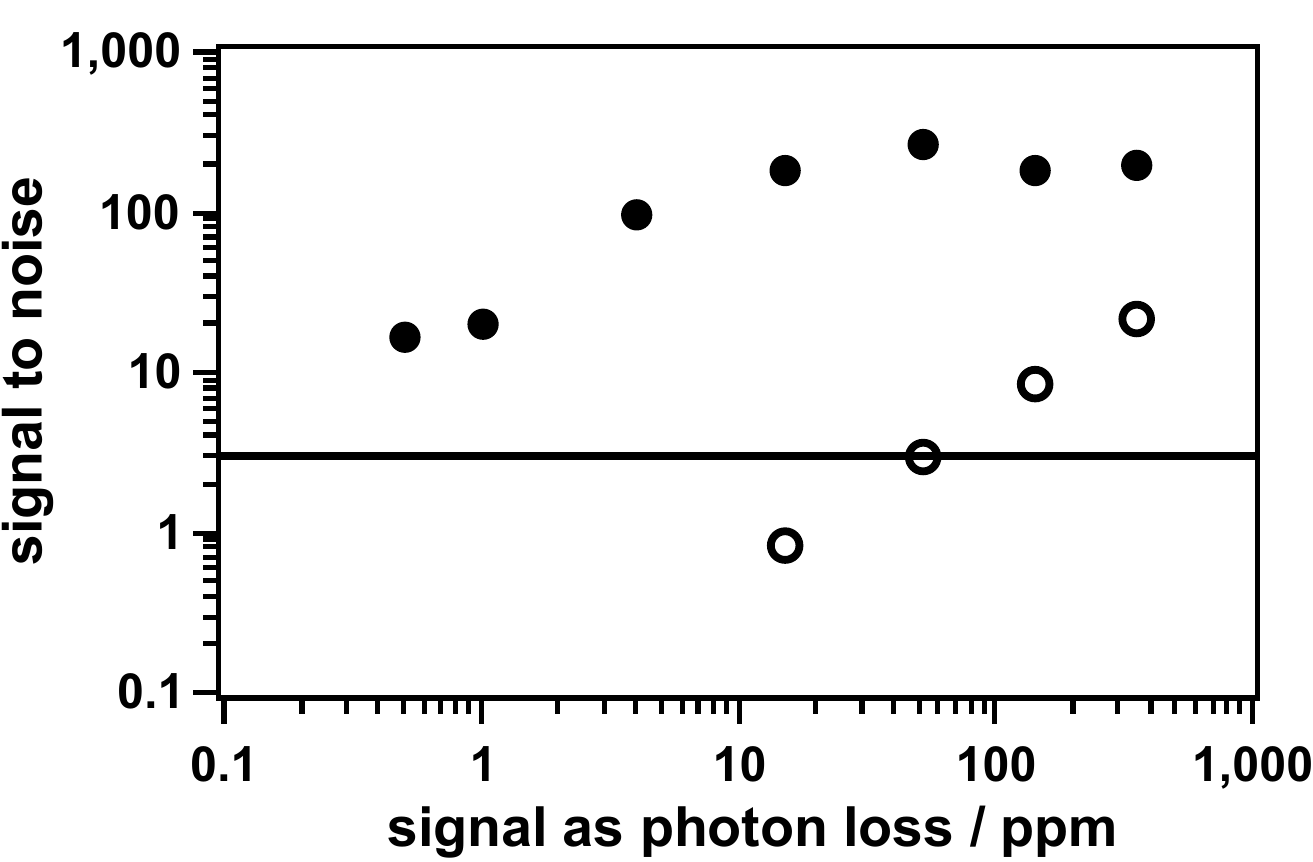}
\caption{Comparison of limits of detection (LOD) for CELIF ($\bullet$) and CRD ($\circ$) measurements of jet-cooled BPEB. The signal-to-noise ratio is the baseline-corrected signal divided by standard deviation of the baseline noise ($\sigma_\text{b}$). The horizontal axis shows the baseline-corrected signal expressed as the photon loss per trip. The typical LOD of $3\sigma_\text{b}$ is indicated by the horizontal line. The data was extracted from simultaneously recorded transients  at 319.69~nm (\emph{cf.}\ fig.\ \ref{fig:BPEBspectra}).}
\label{fig:det-limit}
\end{figure}

In a LIF experiment, the LOD is defined by the baseline noise of the detection system. At very low signals, when the photodetector amplification is high, the baseline noise is amplified as well. The absolute noise typically increases with signal as can be seen in the upper trace of fig.\ \ref{fig:powerdep}b.

We quantified the limits of detection (LOD) of simultaneously recorded CELIF and CRD signals following the recommendations in ref.~\onlinecite{ACS1980}. The gross analyte signal, $S_\text{t}$, was determined from the strongest absorption in fig.\ \ref{fig:BPEBspectra} (319.69~nm). The system blank, $S_\text{b}$, was recorded far off resonance at 320.98~nm and comprises the fluctuation in nozzle intensity, Rayleigh scattering and the noise of the detection system. Data is recorded as a function of sample density and averaged over 2500 laser shots. The noise of the system blank, $\sigma_\text{b}$, is determined by the standard deviation. Figure \ref{fig:det-limit} shows the net signal over the noise, $(S_\text{t} - S_\text{b}) / \sigma_\text{b}$, as a function of the net signal, $S_\text{t} - S_\text{b}$, for both CELIF and CRDS. The commonly accepted limit of detection of $3\sigma_\text{b}$ is indicated by the horizontal line in the plot. Any data point above this line is a detected signal with at least a 99.7\% confidence limit. The CRD measurement at 52~ppm with $(S_\text{t} - S_\text{b}) / \sigma_\text{b}=3$ corresponds to the strongest absorption line in the spectrum fig.\ \ref{fig:BPEBspectra}c. This LOD corresponds to a minimum detectable absorption coefficient of $\alpha_\text{min}=6\cdot 10^{-7}$ cm$^{-1}$.

In the CRD measurement, $\sigma_\text{b}$ is independent of sample concentration leading to the linear dependence seen in fig.\ \ref{fig:det-limit}. As in single-pass LIF or CRD measurements, the CELIF signal increases linearly with sample concentration. Due to the large Rayleigh scattering cross section of BPEB, in these measurements the system blank $S_\text{b}$ is proportional to the sample density. This means that at high concentrations $\sigma_\text{b}$ becomes dominated by the shot-to-shot fluctuations in sample density leading to the flattening of the signal-to-noise ratio in fig.\ \ref{fig:det-limit}. 

In our measurements, the accessible range of sample concentrations is limited by the vapor pressure and the temperature control of the sample oven (low concentrations) and the finite sample volume and the thermal stability of the molecule (high concentrations). Higher sample concentrations than shown in fig.\ \ref{fig:det-limit} could have been detected by both CELIF and CRDS. 
Even if the CRD decay is too short to be measured, the fluorescence can be linearly recorded until saturation caused by optical and sample density sets in. This implies that very strong and weak transitions can be measured in the same scan where the calibration is maintained across the recorded spectrum. 

Our lowest CELIF signal-to-noise value of 17 indicates that considerably lower concentrations are accessible. Based on the fluctuations of the signal blank, that is dominated by BPEB Rayleigh scattering, we estimate the CELIF LOD as 0.1 ppm or $\alpha_\text{min}=1.5\cdot 10^{-9}$ cm$^{-1}$, which constitutes at least a factor of 400 improvement. For our systems, signal levels were sufficient to analyze integrated LIF traces. However, for samples with lower absorbances or fluorescence quantum yields, photon counting can be used to further lower the detection limit (see the following subsection).

We examined the noise in the CRD and CELIF signals using the N$_2$ Rayleigh scattering as a function of N$_2$ pressure in a filled cavity as shown in fig.\ \ref{fig:powerdep}b. For the CRD measurement the noise is independent of the N$_2$ pressure as expected.\cite{Berden2009} Both the noise of the integrated signals on the LIF, $\sigma(S\LIF)$, and the CRD detector, $\sigma(I\CRD)$, contribute to the noise of the CELIF measurement, $\sigma(S\CELIF)$, according to eq.\ \ref{eq:CELIF_norm}. We found that $\sigma(S\LIF)$ increases with signal, according to a typical PMT response, while $\sigma(I\CRD)$ is almost constant, leading overall to the observed increase of $\sigma(S\CELIF)$ with pressure, \emph{i.e.}\ signal.

In a filled cavity at high pressure, the signal-to-noise ratios of the CELIF and CRD measurements are comparable whereas at low pressures the CELIF ratio is two times larger. Compared to this, in the BPEB molecular beam measurements the signal-to-noise ratio and limit of detection are three orders of magnitude better for CELIF. CRD measures the integrated-column density along the cavity axis whereas CELIF, like LIF, images the density in a localised probe volume. In the N$_2$ Rayleigh scattering experiment, the CRD sample length is about 200 times longer than the length of the LIF probe volume. For the CRD measurement, this considerable increase in sample length almost compensates for the higher sensitivity of the CELIF measurement.

In light of this, CELIF is---like LIF---best applied to localized samples, \emph{e.g.}\ found in molecular beams, in flames or at interfaces. The strength of CRD lies in the long effective path length, \emph{e.g.}\ of a filled cavity. In case of a localized sample volume the effective path length through the sample can be reduced by orders of magnitude such that this crucial advantage of CRD is lost, as demonstrated in fig.\ \ref{fig:det-limit}.  Even for a filled cavity measurement, CELIF improves the signal-to-noise ratio in comparison with CRD, particularly at low sample concentrations, as seen in fig.\ \ref{fig:powerdep}b.

\subsection{Absolute quantities from LIF measurements}

So far, we have discussed how, from the CRD measurement, we can directly extract the absorption coefficient, $\alpha$, which in turn can be used to provide an absolute calibration for the CELIF measurement. This requires a measurement range in which both CRD and CELIF measurements provide a non-zero $\alpha$, \emph{cf.}\ eqs\ \ref{eq:alpha_CRD} and \ref{eq:cal_LIF}. As in conventional CRD measurements, for a known sample length the knowledge of either the sample density or the absorption cross section will allow the absolute measurement of the other.  However, in these BPEB measurements using a molecular beam, the sample density cannot easily be determined. We have since measured absolute sample densities of the deuterated mercapto radical (SD) in a dilute molecular beam where the absorption cross section is known.\cite{Mizouri2013} Employing photon counting, the lowest detected sample density was $1.1\cdot 10^{5}$~cm$^{-3}$ corresponding to an $\alpha_\text{min}=7.9\cdot 10^{-11}$~cm$^{-1}$. The SD and BPEB measurements used a similar CELIF setup and detection wavelength. This further improved CELIF LOD is by a factor of 7500 superior to the CRD measurements presented here.  We can therefore conclude that CELIF extends the dynamic range of absorbance measurements compared to a sole CRD measurement by at least three orders of magnitude. 

LIF is a popular technique for quantitative measurements of species in flames due to its high sensitivity, spatial resolution and non-invasive nature. However, for absolute calibration, separate measurements like Rayleigh scattering and, more recently, CRD spectroscopy\cite{Dreyer2001,Luque2004,Lamoureux2010} were used. Variations in optical setup and sample composition between the sequential LIF and CRD measurement may affect the calibration. 
In contrast, our single-beam CELIF method uses the simultaneous and correlated LIF and CRD measurements of the same sample with the same laser pulse to give a robust and consistent calibration.

\subsection{Consideration of fluorescence lifetimes}

\begin{figure}
	\includegraphics[width=8.6cm]{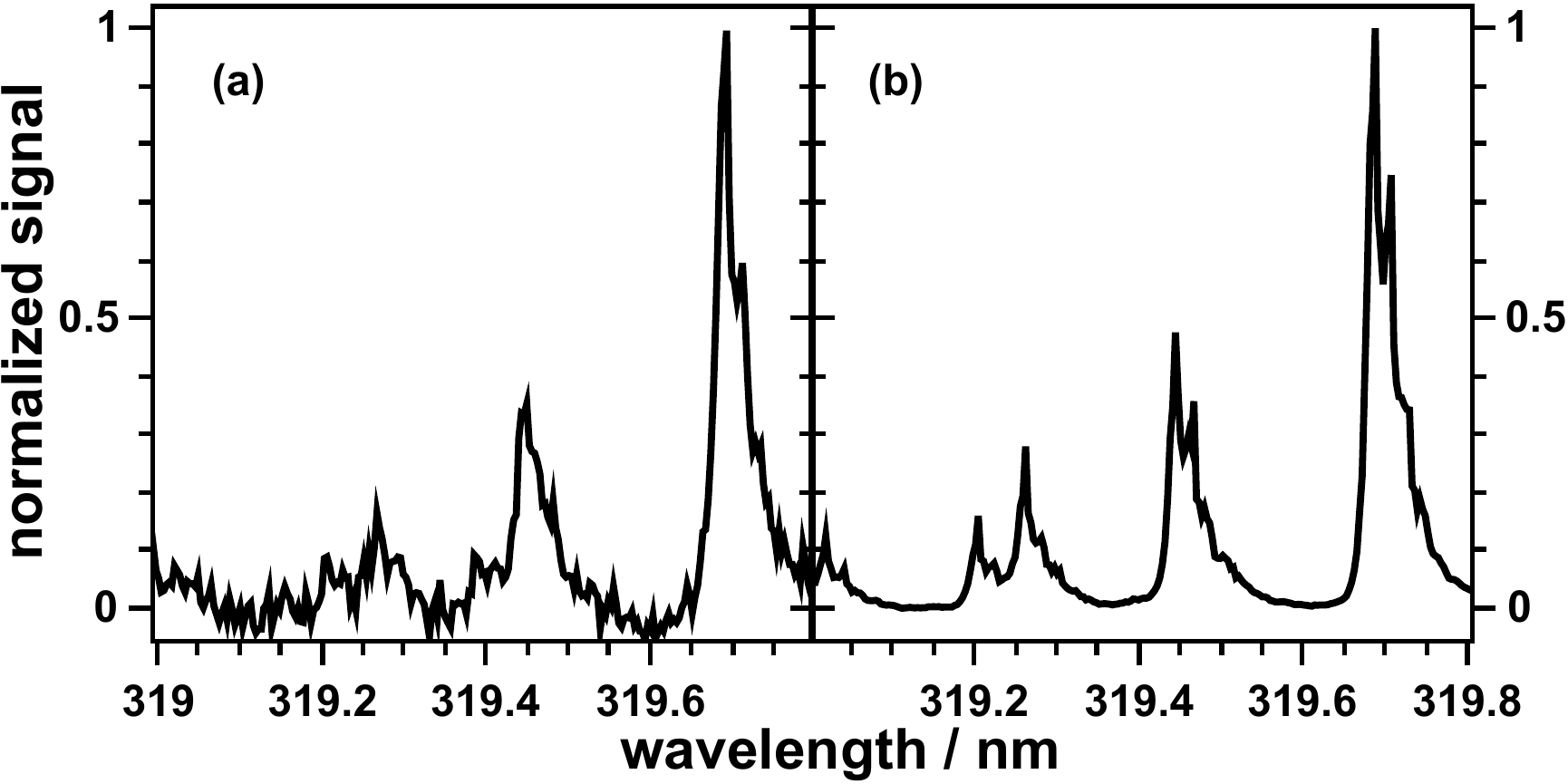}
	\caption{Comparison of normalized BPEB spectra from (a) single-pass LIF and (b) CELIF. The LIF probe volume was approximately 30 times larger than for CELIF and both spectra were recorded with similar sample densities.}
\label{fig:CELIFvLIF}
\end{figure}

The CELIF method is equally applicable to short and long fluorescence lifetimes compared to the laser pulse length.  Short fluorescence lifetimes impose challenges on single-pass, pulsed LIF. The fluorescence signal is obscured by the stray light of the excitation pulse and the Rayleigh scattered light from the sample and cannot easily be discriminated against by gated detection. Stray-light-free signal may only be sampled over a limited temporal range leading to a large noise on the digitized signal. The comparison of  fluorescence excitation spectra of BPEB with a fluorescence lifetime of $\tau_\text{F} \approx 500$~ps\cite{Fujiwara2008} obtained with CELIF and single-pass LIF under similar conditions is shown in fig.\ \ref{fig:CELIFvLIF}. 
For the single-pass LIF measurement, the laser beam was expanded such that the probe volume was approximately 30 times larger than for CELIF in order to increase the signal-to-noise ratio and to limit saturation. The LIF signal was normalized on a shot-to-shot basis against the laser intensity recorded on a pyro detector at the exit window. Stray light was suppressed using a long-pass filter (Semrock, 341 nm blocking edge BrightLine) in front of the LIF photomultiplier. The observed noise on the LIF baseline is  $\sim$25 times larger than in the CELIF spectrum. This demonstrates the difficulty measuring the 500~ps fluorescence free from stray light using a 5~ns laser pulse. As discussed earlier, in CELIF measurements stray light is not supported by the cavity. Although we occasionally observe a small initial peak due to stray light at the very start of the CRD and LIF transients, both these peaks can be completely removed by appropriate gating of the signal. Furthermore, the remaining long LIF signal can be digitized with hundreds of sample points reducing the digitization noise significantly. In order to separate the stray light from fast fluorescence signal, ps lasers and fast signal digitization need to be employed as demonstrated by the fluorescence lifetime measurements of BPEB by Fujiwara \emph{et al.}\cite{Fujiwara2008}

For short fluorescence lifetimes, the LIF transient will follow the CRD transient. In this case, only a part of the transients needs to be integrated, \emph{e.g.}\ using the same limits for both the CRD and LIF signals, to ensure the correct CELIF normalization, see fig.\ \ref{fig:transients}. For long fluorescence lifetimes, the only necessary change to gain valid CELIF spectra is the full integration of both transients. In principle, by deconvoluting the LIF transient by the CRD transient fluorescence lifetimes can be extracted. Long ring-down times reduce the number of photons per unit time in the cavity and, if combined with long fluorescence lifetimes, can lead to signal levels that require photon counting. 

\subsection{Fluorescence yields}\label{sec:yield}

Several groups have used combinations of CRDS with LIF in order to measure quantum yields or quenching rates. 
Spaanjaars \emph{et al.}\ combined a CRD measurement with LIF in a single setup---although not strictly a single beam experiment as the probe laser was split into two beams that crossed the cylindrical burner at different angles---to extract relative predissociation rates of OH in a flame.\cite{Spaanjaars1997} The simultaneous measurement of the absorption via CRD and the fluorescence from similar probe volumes allowed an accurate calibration of the relative predissociation rates and quantum yields. 
Bahrini \emph{et al.}\ measured absorption and fluorescence excitation spectra of CaBr and CaI with a CELIF type setup to obtain relative quantum yields within individual vibrational bands. Unfortunately, it is not clear how the LIF and CRD measurements were calibrated with respect to each other, in particular, as only some spectra were measured concurrently.\cite{Bahrini2006} 
The experiments by Hagemeister \emph{et al.}\ deployed a similar experimental approach to the one described here to measure relative single-vibronic level fluorescence quantum yields of tropolone and tropolone-water clusters.\cite{Hagemeister1999} However, in order to extract accurate relative quantum yields knowledge of the wavelength dependent mirror transmission is required, see eq.\ \ref{eq:ILIF_approx}.

In the following we describe how CELIF can be used to measure absolute fluorescence quantum yields in a self-calibration scheme with Rayleigh scattering. Considering equations \ref{eq:SLIF} and \ref{eq:ILIF_approx}--\ref{eq:alpha_LIF}, the absorption coefficient is
\begin{equation}
	\alpha(\lambda) = \frac{T(\lambda)}{2g\cdot\Gamma(\lambda)}\cdot S\CELIF.
\end{equation}
Performing a Rayleigh scattering measurement (where $\Gamma=1$) with the CELIF setup, the Rayleigh scattering coefficients, $\alpha_\text{R}=\sigma_\text{R}\rho$, can be extracted from the simultaneous CRD measurement or from known Rayleigh cross sections and sample densities. The fraction $\alpha_\text{R}/S\CELIF_\text{R}$ is the calibration factor $\Kcal_\text{R}(\lambda)=T(\lambda)/2g$. The absolute fluorescence quantum yield, $\Gamma_\text{F}$, is then obtained from a subsequent CELIF measurement using the above calibration,
\begin{equation}\label{eq:qy_absolute}
  \Gamma_\text{F}(\lambda) = \Kcal_\text{R}(\lambda)\cdot \frac{S\CELIF_\text{F}(\lambda)}{\alpha\CRD(\lambda)},
\end{equation}
where the absorption coefficient, $\alpha\CRD$, is determined from the ring-down time. Strictly, the above equation neglects the wavelength dependence of $g$ caused by the potentially different response of the detector and the collection optics to the Rayleigh scattered light and the fluorescence spectrum. 

Compared to a CRD measurement, CELIF can be used over a wider range of wavelengths using the same set of cavity mirrors. The sensitivity of CRD cruicially depends on the large effective path length given by the high mirror reflectivity. Following the derivation of eqs \ref{eq:ILIF_approx} and \ref{eq:CELIF_norm}, the CELIF signal is largely independent of the mirror reflectivity and as a consequence the useful wavelength range of the mirrors is extended.  In particular, once the wavelength-dependent calibration factor $K_\text{R}(\lambda)$ is established, the CELIF measurement is independent of the mirror characteristics.

In summary, in a well characterized LIF setup, where the geometric factor of the detection system, $g$, is known, the three variables that determine the LIF signal are the cross section, $\sigma$, the sample density, $\rho$, and the fluorescence quantum yield, $\Gamma$. As outlined above, knowledge of $\alpha=\sigma\rho$ allows the determination of $\Gamma$. Likewise, the sample density can be measured based on a transition for which absorption and fluorescence are known.

\subsection{CELIF implementation}

The benefits of a CELIF setup can easily be gained on an existing CRD experiment. To extend such a setup with LIF detection, the only additions to the system are collection optics, a photodetector and a recording channel. This straightforward and fairly low-cost addition increases the dynamic range of the previous CRD experiment by multiple orders of magnitude. In turn, a new stand-alone LIF setup requires very careful calibration to measure absolute quantities where the challenge remains that most of these calibrations are based on separate measurements where sample and laser beam are similar but not identical. Adding a cavity and CRD detection to the LIF setup provides an \emph{in situ} absolute calibration of the LIF measurement.

\section{Conclusions}\label{sec:conclusions}

We have demonstrated how a conventional, pulsed CRD setup can be extended by fluorescence detection in a straightforward manner that combines the advantages of both the CRD and LIF techniques. The CELIF technique uses the same laser beam and sample in the cavity. Its simultaneous absorption (CRD) and fluorescence (LIF) detection allows a rigorous absolute calibration of the LIF measurement at sample densities that lead to a measurable reduction in ring-down time. This calibration can subsequently be applied across the entire LIF dynamic range. From the calibrated LIF signal, absolute quantities such as the fluorescence quantum yield, absorption cross section and sample density can be extracted. In our experiment, we have shown how the limited dynamic range of the CRD measurement can be extended by at least three orders of magnitude towards lower absorbances, where the change in ring-down time is too small to be detected. 

CELIF very elegantly overcomes two main obstacles of single-pass LIF. The cavity very effectively suppresses stray light allowing detection on the laser excitation wavelength. The probe laser pulse is stretched in time and the resulting decrease in laser intensity reduces saturation. Our measurements show how CELIF improves the signal-to-noise ratio when compared to our single-pass LIF spectrum. We believe that CELIF is most suited for localized sample volumes, such as molecular beams or surfaces, where CRDS cannot fulfill its full potential due to the small absorption path length. 

In our respective research areas, we apply CELIF to molecular spectroscopy in supersonic beams and dynamics at surfaces. Other research fields where absolute spectroscopic quantities such as cross sections and quantum yields are required include astrochemistry, atmospheric chemistry and plasma physics/chemistry. For the more recently evolved research area of cold and ultra-cold molecules, the measurement of absolute densities of molecules in a trap is important. For example, in order to control the outcome of chemical reactions at sub-Kelvin temperatures with external fields, samples densities will need to be increased to $10^{10}$~cm$^{-3}$.\cite{Carr2009} The extended dynamic range of the CELIF technique is perfectly suited to follow the time evolution of absolute trap densities.

We believe that CELIF is an elegant, easy to implement and cost effective way to gain these absolute quantities over a large dynamic range and we hope that it finds applications in many research fields.

\section{Acknowledgements}
We thank Neil Ord, Sian Matthews, Nicholas Andrews and Rebecca Smith for their various contributions throughout the project and Kelvin Appleby for electronic support. We thank EPSRC for the DTA studentships to support SES and ORW, the Royal Society for the University Research Fellowship for NHN, and Durham University for infrastructure support via the HEFCE Science Research Investment Fund.


\begin{thebibliography}{25}%
\makeatletter
\providecommand \@ifxundefined [1]{%
 \@ifx{#1\undefined}
}%
\providecommand \@ifnum [1]{%
 \ifnum #1\expandafter \@firstoftwo
 \else \expandafter \@secondoftwo
 \fi
}%
\providecommand \@ifx [1]{%
 \ifx #1\expandafter \@firstoftwo
 \else \expandafter \@secondoftwo
 \fi
}%
\providecommand \natexlab [1]{#1}%
\providecommand \enquote  [1]{``#1''}%
\providecommand \bibnamefont  [1]{#1}%
\providecommand \bibfnamefont [1]{#1}%
\providecommand \citenamefont [1]{#1}%
\providecommand \href@noop [0]{\@secondoftwo}%
\providecommand \href [0]{\begingroup \@sanitize@url \@href}%
\providecommand \@href[1]{\@@startlink{#1}\@@href}%
\providecommand \@@href[1]{\endgroup#1\@@endlink}%
\providecommand \@sanitize@url [0]{\catcode `\\12\catcode `\$12\catcode
  `\&12\catcode `\#12\catcode `\^12\catcode `\_12\catcode `\%12\relax}%
\providecommand \@@startlink[1]{}%
\providecommand \@@endlink[0]{}%
\providecommand \url  [0]{\begingroup\@sanitize@url \@url }%
\providecommand \@url [1]{\endgroup\@href {#1}{\urlprefix }}%
\providecommand \urlprefix  [0]{URL }%
\providecommand \Eprint [0]{\href }%
\providecommand \doibase [0]{http://dx.doi.org/}%
\providecommand \selectlanguage [0]{\@gobble}%
\providecommand \bibinfo  [0]{\@secondoftwo}%
\providecommand \bibfield  [0]{\@secondoftwo}%
\providecommand \translation [1]{[#1]}%
\providecommand \BibitemOpen [0]{}%
\providecommand \bibitemStop [0]{}%
\providecommand \bibitemNoStop [0]{.\EOS\space}%
\providecommand \EOS [0]{\spacefactor3000\relax}%
\providecommand \BibitemShut  [1]{\csname bibitem#1\endcsname}%
\let\auto@bib@innerbib\@empty
\bibitem [{\citenamefont {Telle}, \citenamefont {{Ure\~na}},\ and\
  \citenamefont {Donovan}(2007)}]{Donovan2007}%
  \BibitemOpen
  \bibfield  {author} {\bibinfo {author} {\bibfnamefont {H.~H.}\ \bibnamefont
  {Telle}}, \bibinfo {author} {\bibfnamefont {A.~G.}\ \bibnamefont
  {{Ure\~na}}}, \ and\ \bibinfo {author} {\bibfnamefont {R.~J.}\ \bibnamefont
  {Donovan}},\ }\href@noop {} {\emph {\bibinfo {title} {Laser Chemistry}}}\
  (\bibinfo  {publisher} {Wiley},\ \bibinfo {year} {2007})\BibitemShut
  {NoStop}%
\bibitem [{\citenamefont {Kinsey}(1977)}]{Kinsey1977}%
  \BibitemOpen
  \bibfield  {author} {\bibinfo {author} {\bibfnamefont {J.~L.}\ \bibnamefont
  {Kinsey}},\ }\href {\doibase 10.1146/annurev.pc.28.100177.002025} {\bibfield
  {journal} {\bibinfo  {journal} {Annu. Rev. Phys. Chem.}\ }\textbf {\bibinfo
  {volume} {28}},\ \bibinfo {pages} {349} (\bibinfo {year} {1977})}\BibitemShut
  {NoStop}%
\bibitem [{\citenamefont {Zare}(2012)}]{Zare2012}%
  \BibitemOpen
  \bibfield  {author} {\bibinfo {author} {\bibfnamefont {R.~N.}\ \bibnamefont
  {Zare}},\ }\href {\doibase 10.1146/annurev-anchem-062011-143148} {\bibfield
  {journal} {\bibinfo  {journal} {Ann. Rev. Anal. Chem.}\ }\textbf {\bibinfo
  {volume} {5}},\ \bibinfo {pages} {1} (\bibinfo {year} {2012})}\BibitemShut
  {NoStop}%
\bibitem [{\citenamefont {Brecher}\ and\ \citenamefont
  {Riseberg}(1976)}]{Brecher1976}%
  \BibitemOpen
  \bibfield  {author} {\bibinfo {author} {\bibfnamefont {C.}~\bibnamefont
  {Brecher}}\ and\ \bibinfo {author} {\bibfnamefont {L.~A.}\ \bibnamefont
  {Riseberg}},\ }\href@noop {} {\bibfield  {journal} {\bibinfo  {journal}
  {Phys. Rev. B}\ }\textbf {\bibinfo {volume} {13}},\ \bibinfo {pages} {81}
  (\bibinfo {year} {1976})}\BibitemShut {NoStop}%
\bibitem [{\citenamefont {Berden}\ and\ \citenamefont
  {Engeln}(2009)}]{Berden2009}%
  \BibitemOpen
  \bibinfo {editor} {\bibfnamefont {G.}~\bibnamefont {Berden}}\ and\ \bibinfo
  {editor} {\bibfnamefont {R.}~\bibnamefont {Engeln}},\ eds.,\ \href@noop {}
  {\emph {\bibinfo {title} {Cavity Ring-down Spectroscopy: Techniques and
  Applications}}}\ (\bibinfo  {publisher} {Wiley-Blackwell},\ \bibinfo {year}
  {2009})\BibitemShut {NoStop}%
\bibitem [{\citenamefont {O'Keefe}\ and\ \citenamefont
  {Deacon}(1988)}]{O'Keefe1988}%
  \BibitemOpen
  \bibfield  {author} {\bibinfo {author} {\bibfnamefont {A.}~\bibnamefont
  {O'Keefe}}\ and\ \bibinfo {author} {\bibfnamefont {D.~A.~G.}\ \bibnamefont
  {Deacon}},\ }\href {\doibase 10.1063/1.1139895} {\bibfield  {journal}
  {\bibinfo  {journal} {Rev. Sci. Instrum.}\ }\textbf {\bibinfo {volume}
  {59}},\ \bibinfo {pages} {2544} (\bibinfo {year} {1988})}\BibitemShut
  {NoStop}%
\bibitem [{\citenamefont {Lehmann}\ and\ \citenamefont
  {Huang}(2008)}]{Lehmann2008}%
  \BibitemOpen
  \bibfield  {author} {\bibinfo {author} {\bibfnamefont {K.~K.}\ \bibnamefont
  {Lehmann}}\ and\ \bibinfo {author} {\bibfnamefont {H.}~\bibnamefont
  {Huang}},\ }\enquote {\bibinfo {title} {Frontiers of molecular
  spectroscopy},}\ \ (\bibinfo  {publisher} {Elsevier},\ \bibinfo {year}
  {2008})\ Chap.\ \bibinfo {chapter} {Optimal Signal Processing in Cavity
  Ring-Down Spectroscopy}, p.\ \bibinfo {pages} {624}\BibitemShut {NoStop}%
\bibitem [{\citenamefont {Spaanjaars}, \citenamefont {terMeulen},\ and\
  \citenamefont {Meijer}(1997)}]{Spaanjaars1997}%
  \BibitemOpen
  \bibfield  {author} {\bibinfo {author} {\bibfnamefont {J.}~\bibnamefont
  {Spaanjaars}}, \bibinfo {author} {\bibfnamefont {J.}~\bibnamefont
  {terMeulen}}, \ and\ \bibinfo {author} {\bibfnamefont {G.}~\bibnamefont
  {Meijer}},\ }\href {\doibase 10.1063/1.474621} {\bibfield  {journal}
  {\bibinfo  {journal} {J. Chem. Phys.}\ }\textbf {\bibinfo {volume} {{107}}},\
  \bibinfo {pages} {2242} (\bibinfo {year} {{1997}})}\BibitemShut {NoStop}%
\bibitem [{\citenamefont {Hagemeister}\ \emph {et~al.}(1999)\citenamefont
  {Hagemeister}, \citenamefont {Arrington}, \citenamefont {Giles},
  \citenamefont {Quimpo}, \citenamefont {Zhang},\ and\ \citenamefont
  {Zwier}}]{Hagemeister1999}%
  \BibitemOpen
  \bibfield  {author} {\bibinfo {author} {\bibfnamefont {F.~C.}\ \bibnamefont
  {Hagemeister}}, \bibinfo {author} {\bibfnamefont {C.~A.}\ \bibnamefont
  {Arrington}}, \bibinfo {author} {\bibfnamefont {B.~J.}\ \bibnamefont
  {Giles}}, \bibinfo {author} {\bibfnamefont {B.}~\bibnamefont {Quimpo}},
  \bibinfo {author} {\bibfnamefont {L.}~\bibnamefont {Zhang}}, \ and\ \bibinfo
  {author} {\bibfnamefont {T.~S.}\ \bibnamefont {Zwier}},\ }\enquote {\bibinfo
  {title} {Cavity-ringdown spectroscopy},}\ \ (\bibinfo  {publisher} {ACS},\
  \bibinfo {year} {1999})\ pp.\ \bibinfo {pages} {210--232}\BibitemShut
  {NoStop}%
\bibitem [{\citenamefont {Bahrini}\ \emph {et~al.}(2006)\citenamefont
  {Bahrini}, \citenamefont {Douin}, \citenamefont {Rostas},\ and\ \citenamefont
  {Taieb}}]{Bahrini2006}%
  \BibitemOpen
  \bibfield  {author} {\bibinfo {author} {\bibfnamefont {C.}~\bibnamefont
  {Bahrini}}, \bibinfo {author} {\bibfnamefont {S.}~\bibnamefont {Douin}},
  \bibinfo {author} {\bibfnamefont {J.}~\bibnamefont {Rostas}}, \ and\ \bibinfo
  {author} {\bibfnamefont {G.}~\bibnamefont {Taieb}},\ }\href {\doibase
  10.1016/j.cplett.2006.09.073} {\bibfield  {journal} {\bibinfo  {journal}
  {Chem. Phys. Lett.}\ }\textbf {\bibinfo {volume} {{432}}},\ \bibinfo {pages}
  {1} (\bibinfo {year} {{2006}})}\BibitemShut {NoStop}%
\bibitem [{\citenamefont {Tokaryk}, \citenamefont {Adam},\ and\ \citenamefont
  {Slaney}(2007)}]{Tokaryk2007}%
  \BibitemOpen
  \bibfield  {author} {\bibinfo {author} {\bibfnamefont {D.}~\bibnamefont
  {Tokaryk}}, \bibinfo {author} {\bibfnamefont {A.}~\bibnamefont {Adam}}, \
  and\ \bibinfo {author} {\bibfnamefont {M.}~\bibnamefont {Slaney}},\ }\href
  {\doibase 10.1016/j.cplett.2006.11.053} {\bibfield  {journal} {\bibinfo
  {journal} {Chem. Phys. Lett.}\ }\textbf {\bibinfo {volume} {433}},\ \bibinfo
  {pages} {264} (\bibinfo {year} {2007})}\BibitemShut {NoStop}%
\bibitem [{\citenamefont {Richman}\ \emph {et~al.}(2005)\citenamefont
  {Richman}, \citenamefont {Kachanov}, \citenamefont {Paldus},\ and\
  \citenamefont {Strawa}}]{Richman2005}%
  \BibitemOpen
  \bibfield  {author} {\bibinfo {author} {\bibfnamefont {B.~A.}\ \bibnamefont
  {Richman}}, \bibinfo {author} {\bibfnamefont {A.~A.}\ \bibnamefont
  {Kachanov}}, \bibinfo {author} {\bibfnamefont {B.~A.}\ \bibnamefont
  {Paldus}}, \ and\ \bibinfo {author} {\bibfnamefont {A.~W.}\ \bibnamefont
  {Strawa}},\ }\href {\doibase 10.1364/OPEX.13.003376} {\bibfield  {journal}
  {\bibinfo  {journal} {Optics Express}\ }\textbf {\bibinfo {volume} {13}},\
  \bibinfo {pages} {3376} (\bibinfo {year} {2005})}\BibitemShut {NoStop}%
\bibitem [{\citenamefont {Dreyer}, \citenamefont {Spuler},\ and\ \citenamefont
  {Linne}(2001)}]{Dreyer2001}%
  \BibitemOpen
  \bibfield  {author} {\bibinfo {author} {\bibfnamefont {C.~B.}\ \bibnamefont
  {Dreyer}}, \bibinfo {author} {\bibfnamefont {S.~M.}\ \bibnamefont {Spuler}},
  \ and\ \bibinfo {author} {\bibfnamefont {M.}~\bibnamefont {Linne}},\ }\href
  {\doibase 10.1080/00102200108907863} {\bibfield  {journal} {\bibinfo
  {journal} {Combust. Sci. Technol.}\ }\textbf {\bibinfo {volume} {171}},\
  \bibinfo {pages} {163} (\bibinfo {year} {2001})}\BibitemShut {NoStop}%
\bibitem [{\citenamefont {Luque}\ \emph {et~al.}(2004)\citenamefont {Luque},
  \citenamefont {Berg}, \citenamefont {Jeffries}, \citenamefont {Smith},
  \citenamefont {Crosley},\ and\ \citenamefont {Scherer}}]{Luque2004}%
  \BibitemOpen
  \bibfield  {author} {\bibinfo {author} {\bibfnamefont {J.}~\bibnamefont
  {Luque}}, \bibinfo {author} {\bibfnamefont {P.}~\bibnamefont {Berg}},
  \bibinfo {author} {\bibfnamefont {J.}~\bibnamefont {Jeffries}}, \bibinfo
  {author} {\bibfnamefont {G.}~\bibnamefont {Smith}}, \bibinfo {author}
  {\bibfnamefont {D.}~\bibnamefont {Crosley}}, \ and\ \bibinfo {author}
  {\bibfnamefont {J.}~\bibnamefont {Scherer}},\ }\href {\doibase
  10.1007/s00340-003-1331-3} {\bibfield  {journal} {\bibinfo  {journal} {Appl.
  Phys. B}\ }\textbf {\bibinfo {volume} {78}},\ \bibinfo {pages} {93} (\bibinfo
  {year} {2004})}\BibitemShut {NoStop}%
\bibitem [{\citenamefont {Lamoureux}\ \emph {et~al.}(2010)\citenamefont
  {Lamoureux}, \citenamefont {Desgroux}, \citenamefont {Bakali},\ and\
  \citenamefont {Pauwels}}]{Lamoureux2010}%
  \BibitemOpen
  \bibfield  {author} {\bibinfo {author} {\bibfnamefont {N.}~\bibnamefont
  {Lamoureux}}, \bibinfo {author} {\bibfnamefont {P.}~\bibnamefont {Desgroux}},
  \bibinfo {author} {\bibfnamefont {A.~E.}\ \bibnamefont {Bakali}}, \ and\
  \bibinfo {author} {\bibfnamefont {J.}~\bibnamefont {Pauwels}},\ }\href
  {\doibase 10.1016/j.combustflame.2010.03.013} {\bibfield  {journal} {\bibinfo
   {journal} {Combust Flame}\ }\textbf {\bibinfo {volume} {157}},\ \bibinfo
  {pages} {1929 } (\bibinfo {year} {2010})}\BibitemShut {NoStop}%
\bibitem [{\citenamefont {Greaves}\ \emph {et~al.}(2006)\citenamefont
  {Greaves}, \citenamefont {Flynn}, \citenamefont {Futcher}, \citenamefont
  {Wrede}, \citenamefont {Lydon}, \citenamefont {Low}, \citenamefont {Rutter},\
  and\ \citenamefont {Beeby}}]{Greaves2006}%
  \BibitemOpen
  \bibfield  {author} {\bibinfo {author} {\bibfnamefont {S.~J.}\ \bibnamefont
  {Greaves}}, \bibinfo {author} {\bibfnamefont {E.~L.}\ \bibnamefont {Flynn}},
  \bibinfo {author} {\bibfnamefont {E.~L.}\ \bibnamefont {Futcher}}, \bibinfo
  {author} {\bibfnamefont {E.}~\bibnamefont {Wrede}}, \bibinfo {author}
  {\bibfnamefont {D.~P.}\ \bibnamefont {Lydon}}, \bibinfo {author}
  {\bibfnamefont {P.~J.}\ \bibnamefont {Low}}, \bibinfo {author} {\bibfnamefont
  {S.~R.}\ \bibnamefont {Rutter}}, \ and\ \bibinfo {author} {\bibfnamefont
  {A.}~\bibnamefont {Beeby}},\ }\href {\doibase 10.1021/jp054426h} {\bibfield
  {journal} {\bibinfo  {journal} {J. Phys, Chem. A}\ }\textbf {\bibinfo
  {volume} {110}},\ \bibinfo {pages} {2114} (\bibinfo {year}
  {2006})}\BibitemShut {NoStop}%
\bibitem [{\citenamefont {Beeby}\ \emph {et~al.}(2002)\citenamefont {Beeby},
  \citenamefont {Findlay}, \citenamefont {Low},\ and\ \citenamefont
  {Marder}}]{Beeby2002}%
  \BibitemOpen
  \bibfield  {author} {\bibinfo {author} {\bibfnamefont {A.}~\bibnamefont
  {Beeby}}, \bibinfo {author} {\bibfnamefont {K.}~\bibnamefont {Findlay}},
  \bibinfo {author} {\bibfnamefont {P.~J.}\ \bibnamefont {Low}}, \ and\
  \bibinfo {author} {\bibfnamefont {T.~B.}\ \bibnamefont {Marder}},\
  }\href@noop {} {\bibfield  {journal} {\bibinfo  {journal} {J. Am. Chem.
  Soc.}\ }\textbf {\bibinfo {volume} {124}},\ \bibinfo {pages} {8280} (\bibinfo
  {year} {2002})}\BibitemShut {NoStop}%
\bibitem [{Cro()}]{CrossSectionNote}%
  \BibitemOpen
  \href@noop {} {}\bibinfo {note} {We estimate an absorption cross section of
  the strongest vibronic line at 319.69~nm in excess of $5\cdot
  10^{-15}$~cm$^2$ from the gas-phase UV spectrum of BPEB at 120~$^\circ$C
  (ref.\ \onlinecite{Greaves2006}). The maximum of the gas-phase spectrum was
  scaled to match the maximum extinction coefficient of BPEB in cyclohexane
  solution at room temperature: 58\,000~dm$^3$~mol$^{-1}$~cm$^{-1}$ (ref.\
  \onlinecite{Beeby2002}). We assume that the cooling in the supersonic
  expansion increases the population in the vibrational ground state by at
  least a factor of 100.}\BibitemShut {Stop}%
\bibitem [{\citenamefont {Fujiwara}, \citenamefont {Zgierski},\ and\
  \citenamefont {Lim}(2008)}]{Fujiwara2008}%
  \BibitemOpen
  \bibfield  {author} {\bibinfo {author} {\bibfnamefont {T.}~\bibnamefont
  {Fujiwara}}, \bibinfo {author} {\bibfnamefont {M.~Z.}\ \bibnamefont
  {Zgierski}}, \ and\ \bibinfo {author} {\bibfnamefont {E.~C.}\ \bibnamefont
  {Lim}},\ }\href {\doibase 10.1021/jp711064g} {\bibfield  {journal} {\bibinfo
  {journal} {J. Phys, Chem. A}\ }\textbf {\bibinfo {volume} {112}},\ \bibinfo
  {pages} {4736} (\bibinfo {year} {2008})}\BibitemShut {NoStop}%
\bibitem [{\citenamefont {Zalicki}\ and\ \citenamefont
  {Zare}(1995)}]{Zalicki1995}%
  \BibitemOpen
  \bibfield  {author} {\bibinfo {author} {\bibfnamefont {P.}~\bibnamefont
  {Zalicki}}\ and\ \bibinfo {author} {\bibfnamefont {R.~N.}\ \bibnamefont
  {Zare}},\ }\href {\doibase 10.1063/1.468647} {\bibfield  {journal} {\bibinfo
  {journal} {J. Chem. Phys.}\ }\textbf {\bibinfo {volume} {102}},\ \bibinfo
  {pages} {2708} (\bibinfo {year} {1995})}\BibitemShut {NoStop}%
\bibitem [{\citenamefont {Brown}, \citenamefont {Ravishankara},\ and\
  \citenamefont {Stark}(2000)}]{Brown2000a}%
  \BibitemOpen
  \bibfield  {author} {\bibinfo {author} {\bibfnamefont {S.}~\bibnamefont
  {Brown}}, \bibinfo {author} {\bibfnamefont {A.}~\bibnamefont {Ravishankara}},
  \ and\ \bibinfo {author} {\bibfnamefont {H.}~\bibnamefont {Stark}},\ }\href
  {\doibase 10.1021/jp0013715} {\bibfield  {journal} {\bibinfo  {journal} {J.
  Phys. Chem. A}\ }\textbf {\bibinfo {volume} {{104}}},\ \bibinfo {pages}
  {7044} (\bibinfo {year} {{2000}})}\BibitemShut {NoStop}%
\bibitem [{\citenamefont {Wang}\ and\ \citenamefont {Zhang}(2000)}]{Wang2000}%
  \BibitemOpen
  \bibfield  {author} {\bibinfo {author} {\bibfnamefont {L.}~\bibnamefont
  {Wang}}\ and\ \bibinfo {author} {\bibfnamefont {J.}~\bibnamefont {Zhang}},\
  }\href {\doibase 10.1021/es0011055} {\bibfield  {journal} {\bibinfo
  {journal} {Environ. Sci. Technol.}\ }\textbf {\bibinfo {volume} {34}},\
  \bibinfo {pages} {4221} (\bibinfo {year} {2000})}\BibitemShut {NoStop}%
\bibitem [{\citenamefont {MacDougall}\ and\ \citenamefont
  {Crummett}(1980)}]{ACS1980}%
  \BibitemOpen
  \bibfield  {author} {\bibinfo {author} {\bibfnamefont {D.}~\bibnamefont
  {MacDougall}}\ and\ \bibinfo {author} {\bibfnamefont {W.}~\bibnamefont
  {Crummett}},\ }\href {\doibase 10.1021/ac50064a004} {\bibfield  {journal}
  {\bibinfo  {journal} {Anal. Chem.}\ }\textbf {\bibinfo {volume} {52}},\
  \bibinfo {pages} {2242} (\bibinfo {year} {1980})}\BibitemShut {NoStop}%
\bibitem [{\citenamefont {Mizouri}\ \emph {et~al.}(2013)\citenamefont
  {Mizouri}, \citenamefont {Deng}, \citenamefont {Eardley}, \citenamefont
  {Nahler}, \citenamefont {Wrede},\ and\ \citenamefont {Carty}}]{Mizouri2013}%
  \BibitemOpen
  \bibfield  {author} {\bibinfo {author} {\bibfnamefont {A.}~\bibnamefont
  {Mizouri}}, \bibinfo {author} {\bibfnamefont {L.~Z.}\ \bibnamefont {Deng}},
  \bibinfo {author} {\bibfnamefont {J.~S.}\ \bibnamefont {Eardley}}, \bibinfo
  {author} {\bibfnamefont {N.~H.}\ \bibnamefont {Nahler}}, \bibinfo {author}
  {\bibfnamefont {E.}~\bibnamefont {Wrede}}, \ and\ \bibinfo {author}
  {\bibfnamefont {D.}~\bibnamefont {Carty}},\ }\href@noop {} {\bibfield
  {journal} {\bibinfo  {journal} {Phys. Chem. Chem. Phys.}\ } (\bibinfo {year}
  {2013})},\ \bibinfo {note} {accepted, arXiv:1308.2105v2}\BibitemShut
  {NoStop}%
\bibitem [{\citenamefont {Carr}\ \emph {et~al.}(2009)\citenamefont {Carr},
  \citenamefont {DeMille}, \citenamefont {Krems},\ and\ \citenamefont
  {Ye}}]{Carr2009}%
  \BibitemOpen
  \bibfield  {author} {\bibinfo {author} {\bibfnamefont {L.~D.}\ \bibnamefont
  {Carr}}, \bibinfo {author} {\bibfnamefont {D.}~\bibnamefont {DeMille}},
  \bibinfo {author} {\bibfnamefont {R.~V.}\ \bibnamefont {Krems}}, \ and\
  \bibinfo {author} {\bibfnamefont {J.}~\bibnamefont {Ye}},\ }\href@noop {}
  {\bibfield  {journal} {\bibinfo  {journal} {New J. Phys.}\ }\textbf {\bibinfo
  {volume} {11}},\ \bibinfo {pages} {055049} (\bibinfo {year}
  {2009})}\BibitemShut {NoStop}%
\end{thebibliography}
\end{document}